\documentclass[useAMS,usenatbib,usegraphicx]{mn2e}

\usepackage[dvips]{graphics}
\usepackage{dcolumn}% Align table columns on decimal point
\usepackage{bm}% bold math
\usepackage{amssymb,amsmath}
\usepackage{epsfig}
\usepackage{color}
\usepackage{amsfonts}
\usepackage{amscd}
\usepackage{dsfont}

\title[21-cm Fluctuation PDFs]{Statistics of 21-cm Fluctuations in 
Cosmic Reionization Simulations: PDFs and Difference PDFs}
\author[Gluscevic and Barkana]{Vera Gluscevic$^1$ and
Rennan Barkana$^{1,2}$\thanks{E-mail: vera@astro.caltech.edu (VG);
barkana@wise.tau.ac.il (RB).}\\ $^1$ Division of Physics, Mathematics
and Astronomy, California Institute of Technology, Mail Code 350-17,
Pasadena, CA 91125, USA \\ $^2$ Raymond and Beverly Sackler School of
Physics and Astronomy, Tel Aviv University, Tel Aviv 69978, Israel}

\begin{document}

\pagerange{\pageref{firstpage}--\pageref{lastpage}} \pubyear{2009}
\maketitle
\label{firstpage}
\begin{abstract}
In the coming decade, low-frequency radio arrays will begin to probe
the epoch of reionization via the redshifted 21-cm hydrogen line. Successful
interpretation of these observations will require effective
statistical techniques for analyzing the data. Due to the difficulty
of these measurements, it is important to develop techniques
beyond the standard power spectrum
analysis in order to offer independent confirmation of
the reionization history, probe different aspects of the topology of
reionization, and have different systematic errors. In order to assess
the promise of probability distribution functions (PDFs) as
statistical analysis tools in 21-cm cosmology, we first measure the
21-cm brightness temperature (one-point) PDFs in six different
reionization simulations. We then parametrize their most distinct
features by fitting them to a simple model. Using the same
simulations, we also present the first measurements of difference PDFs
in simulations of reionization. We find that while these statistics
probe the properties of the ionizing sources, they are relatively
independent of small-scale, sub-grid astrophysics. We discuss the
additional information that the difference PDF can provide on top of
the power spectrum and the one-point PDF.
\end{abstract}


\section{Introduction: Cosmic Reionization and 21-cm Cosmology}\label{intro}

The first stars and quasars reionized the intergalactic medium (IGM)
during the epoch of reionization (EOR) \citep{Review}. From the
Ly$\alpha$ absorption in the spectra of distant quasars, we know that
the EOR ended at $z\sim 6$ \citep[see, e.g., ][]{dj05}, and from the
cosmic microwave background (CMB) polarization maps we can infer that
it started no later than $z\sim 15$
\citep{WMAP}. Understanding how the reionization process developed
over time offers a way to answer some of the critical questions of
modern cosmology concerning properties of the first sources of light
in the universe. Among the most promising observational probes of the
EOR is the 21-cm spectral line, from hyperfine splitting in the ground
state of hydrogen, with an energy of $5.9\times 10^{-6}$eV that
corresponds to the rest-frame frequency of 1420 MHz. The redshifted
21-cm emission from neutral regions of the IGM during reionization is
estimated to be a $1\%$ correction to the energy density of the
CMB. It is expected to display angular structure and frequency
structure, due to the inhomogeneities in the gas density, ionized
fraction, $x_i$\footnote{The ionized fraction $x_i$ is the ratio of
the number of protons to the total number of hydrogen atoms.}, and
spin temperature \citep{md97} of the emitting gas.

Statistical detection of the large-scale brightness fluctuations is
within the scope of a number of experiments that are presently being
built, such as the Murchison Widefield Array (MWA) and the Low
Frequency Array (LOFAR)\citep[for reviews see, e.g.,][]{fob06,
bl07}. In this context, it is important to develop appropriate
statistical tools to be employed in analyzing the incoming data. Such
development is facilitated by the fact that the N-body and
radiative-transfer simulations of reionization have begun to reach the
large scales of order 100 comoving Mpc \citep{mq07,iliev,santos}
needed to capture the evolution of the IGM during the EOR. These
simulated data cubes can be used to test various statistical tools
proposed for extracting information about the properties of the IGM
during reionization.

So far, studies of the statistics of the 21-cm fluctuations have
mainly focused on the power spectrum of the brightness temperature,
$T_b$ \citep{miguel, CfA}. While this statistic is fully
representative at the onset of the EOR, where the Gaussian primordial
density fluctuations drive the 21-cm fluctuations, it ceases to be so
at later times. Namely, as the reionization process advances, the
mapping between the hydrogen density and $T_b$ becomes highly
non-linear (as evidenced for instance by the bounded domain, $x_i \in
[0,1]$), which results in non-Gaussianity of the probability
distribution function (PDF) of $T_b$.  For this reason, various
authors have started exploring alternative and complementary
statistics \citep[e.g., ][]{fzh04, Ali}, in particular the PDFs and
difference PDFs of the 21-cm brightness temperature \citep{dif08,
ich09}. We test these two statistics on six different simulated data
cubes from
\citet{mq07}. These data cubes are results of different astrophysical inputs 
that produce various reionization histories, all of which are allowed
by the current observational constraints. We measure the one-point
PDFs and difference PDFs and analyze their properties.

The plan of the paper is as follows: in Section~\ref{sims} we briefly
describe the simulation runs used in this paper. In
Section~\ref{onept} we then present the measured one-point PDFs along
with the best fits of the model proposed by \citet{ich09}, and discuss
the main parameters driving the PDF shape. We next present in
Section~\ref{diff} the first measurements of difference PDFs for the
same set of simulations and analyze their properties. We conclude in
Section~\ref{conclude}.

\section{Statistics of 21-cm Fluctuations}\label{stat}

%%%%%%%%%%%%%%%%%%%%%%%%%%%%%%%%%%%%%%%%%%%%%%%%%%%%%%%%%%%%%%%%%%%%
\subsection{Simulations}\label{sims}

In order to interpret future observations of the high-redshift
universe, we need to understand the morphology of HII regions during
reionization, in particular their size distribution and how it is affected by the properties of the ionizing
sources, gas clumping and source suppression from photoheating
feedback. For this
purpose, \citet{mq07} ran a $1024^3$ N-body simulation in a box of
size $65.6h^{-1} \simeq 94$ Mpc to model the density field,
post-processing it using a suite of radiative transfer
simulations. The authors assumed a standard $\Lambda$CDM cosmology,
with $n_s = 1$, $\sigma _8 = 0.9$, $\Omega_m = 0.3$, $\Omega_\Lambda =
0.7$, $\Omega_b = 0.04$ and $h = 0.7$. The outputs are stored at $50$
million year intervals, roughly between redshifts 6 and 16. 

The radiative transfer code assumes sharp HII fronts, which are traced
at subgrid scales. The properties of the sources are chosen in most
cases so that reionization ends near z $\sim$ $7$. A soft ultraviolet
spectrum that scales as $\nu ^4$ is assumed for each source. The
typical luminosity of a halo of mass $m$ is taken to be $\dot N(m) = 3
\times {10^{49}}m/({10^8}{M_ \odot })$ ionizing photons per second.
This corresponds to a halo star formation rate of $\dot S(m) = f_{\rm
esc}^{-1}\, m/({10^{10}}{M_\odot })$ in units of $M_ \odot\, {\rm
yr}^{-1}$, for an escape fraction of $f_{\rm esc}$ and a Salpeter
IMF. The N-body simulation resolves haloes down to ${10^9}{M_\odot }$,
but since the effect of smaller mass haloes cannot be neglected, the
effect of haloes down to ${10^8}{M_ \odot }$ is included in some of the
runs with a merger tree (see Table~\ref{simulations}).

%%%%%%%%%%%%%%%%%%%%%%%%%%%%%%%%%%%%%%%%%%%%%%%%%%%%%%%%%%%%%%%%%%%%%%%%%
%%%%%%%%%%%%%%%%%%%%%%                     TABLE 1
\begin{table*}
\centering
\begin{tabular}{c c c c}
\hline \hline
Simulation & Merger tree haloes & ${\dot N}$ [photons sec$^{-1}$] & Comments\\
\hline 

%$S0$ & ? & $?$\\ \\ 
$S1$ & Yes & $2\times 10^{49}M_8$ & -\\ \\
$S4$ & No & $C_{S4}M_8$ & Includes only haloes with 
$m>4\times 10^{10}M_{\odot}$\\ \\
$C5$ & No & $6\times 10^{49}M_8$ & Structure on small scales; 
$C_{\text{cell}}=4+3\delta _{\text{C}}$\\ \\
$F2$ & Yes & $2\times 10^{49}M_8$ & Includes feedback on $m<M_J/2$; 
$\tau_{ SF}=20\,\text{Myr}$\\ \\
$M2$ & No & $9\times 10^{49}M_8$ & Includes minihaloes with 
$m_{\text{mini}}>10^5 M_{\odot}$\\ \\
$Z1$ & Yes & $1\times 10^{50}M_8$ & Higher source efficiency 
(early reionization)\\

\hline
\\
\end{tabular}
\caption{Details on the radiative transfer simulations from
\citet{mq07}. Merger tree haloes: 'Yes' means that the halo resolution
is supplemented with a merger tree down to $10^8M_{\odot}$. $C_{S4}$
is calibrated such that there is the same output of ionizing photons
in each time-step as in S1. $M_8$ denotes the halo mass in units of
$10^8 M_{\odot}$. $C_{\text{cell}}$ is the subgrid clumping factor,
and $\delta _{\text{C}}$ is the baryonic overdensity smoothed on the
cell scale. $\tau _{SF}$ is the time-scale over which the cool gas in
the source is converted into stars, and $M_J$ is the linear-theory
Jeans mass.\label{simulations}}
\end{table*} 

For the purpose of measuring PDFs and difference PDFs, we choose six
runs, labeled as in \citet{mq07}: S1, S4, C5, F2, M2 and Z1. These runs
differ by the efficiency of the sources and by the halo-mass
resolution. Some runs include feedback from photo-heating, which
suppresses source formation within ionized regions. Others investigate
the impact of clumping, i.e., IGM density inhomogeneities, and include
a subgrid clumping factor $C_{\text{cell}}$ different from
unity. Finally, some runs account for the presence of minihaloes,
which are dense absorbers for ionizing photons and thus tend to extend
the process of reionization. A summary of the parameters of each of
the six runs is presented in Table~\ref{simulations}. The list of the
redshift slices for each data cube is shown in Table~\ref{slices}. For
more details about the simulations, see \citet{mq07}.

%%%%%%%%%%%%%%%%%%%%%%%%%%%%%%%%%%%%%%%%%%%%%%%%%%%%%%%%%%%%%%%%%%%%%%%
%%%%%%%%%%%%%%%%%%%%          TABLE 2
\begin{table*}
\centering
\begin{tabular}{c c c c}
\hline \hline
Simulation & Redshift slices \\
\hline 
%$S0$ & $6.6, 6.9, 7.3, 7.7, 8.2, 8.7, 9.4, 10.1, 11.1, 12.3, 13.9$ \\ \\
$S1$ & \hspace{1.2cm}$6.9, 7.3, 7.7, 8.2, 8.7, 9.4, 10.1, 11.1, 12.3, 
13.9, 16.2$ \\ \\
$S4$ & \hspace{1.8cm}$7.3, 7.7, 8.2, 8.7, 9.4, 10.1, 11.1, 12.3, 
13.9, 16.2$ \\ \\
$C5$ & \hspace{.7cm}$6.6, 6.9, 7.3, 7.7, 8.2, 8.7, 9.4, 10.1, 11.1, 12.3, 
13.9, 16.2$ \\ \\
$F2$ & \hspace{.2cm}$6.3, 6.6, 6.9, 7.3, 7.7, 8.2, 8.7, 9.4, 10.1, 11.1, 
12.3, 13.9, 16.2$ \\ \\
$M2$ & \hspace{1.25cm}$6.9, 7.3, 7.7, 8.2, 8.7, 9.4, 10.1, 11.1, 12.3, 
13.9, 16.2$ \\ \\
$Z1$ & \hspace{4.4cm}$10.1, 11.1, 12.3, 13.9, 16.2$ \\
\hline
\\
\end{tabular}
\caption{List of the simulation runs and corresponding redshift slices 
discussed in this paper. The redshift outputs for each simulation are
spaced in 50 Myr intervals.\label{slices}}
\end{table*} 

%%%%%%%%%%%%%%%%%%%%%%%%%%%%%%%%%%%%%%%%%%%%%%%%%%%%%%%%%%%%%%%%%%%%%%%%

\subsection{One-point PDFs}\label{onept}

\begin{figure}
\includegraphics[width=9cm,keepaspectratio=true]{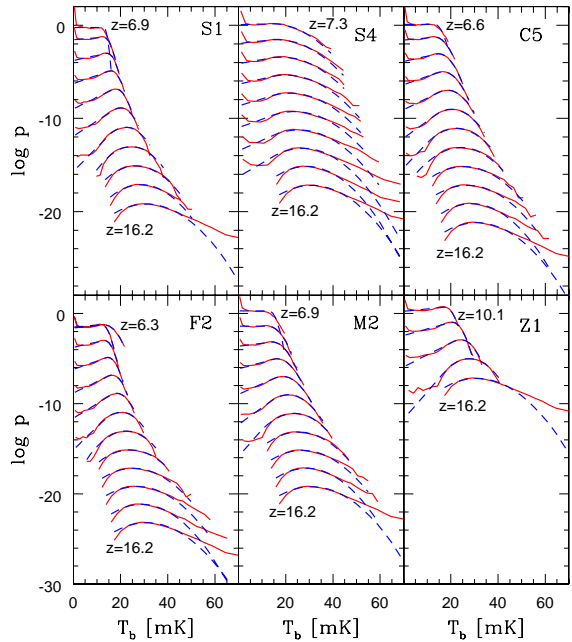}
\caption{One-point PDFs for six different simulation runs, at the 
redshifts listed in Table~\ref{slices}. The $y$-axis is the logarithm
of $p(T_b)$, where $p$ represents the number fraction of points (i.e.,
pixels, or data-cube cells) at a given $T_b$, normalized by the
temperature-bin size. The units of $p$ are mK$^{-1}$. There are $20$
evenly spaced temperature bins in each curve. The size of a bin at
each redshift is set by the temperature range of the cells at that
redshift. The top PDF in each plot is at the lowest redshift of the
data cube, and the redshift increases downward. The $i$-th PDF in each
plot is shifted down the vertical axis by $2(i-1)$ in logarithmic
space, for clarity. We show the measured PDFs (solid red curves) as
well as the best fits of the GED model (dashed blue curves).}
\label{1pt}%
\end{figure}

We measure the one-point PDFs of the observed brightness temperature
$T_b$ of the redshifted 21-cm emission, in the six simulation runs
from \citet{mq07}. We measure $T_b$ in a $32^3$ grid, i.e., $T_b$ is
averaged over cells of size of 2.9 comoving Mpc, at each of the
redshifts listed in Table~\ref{slices}. Note that this scale is much
larger than the simulation's spatial resolution; it is chosen by
balancing the requirement to be large enough to fall close to the
general range of the upcoming observations (corresponding to a
resolution of arcminutes), with the need to be small enough compared
to the simulation box to give a reasonable statistical sample. The
PDFs are shown in Figure~\ref{1pt}.
%~~~~~~~~~~~~~~~~~~~~~~~~~~~~~~~~~~~~~~~~~~~~~~~~~~~~~~~~~~~~~~~~~~~~~
%                                        FIGURE 1

As seen in Figure~\ref{1pt}, the one-point PDFs are Gaussian-like at the
highest redshifts and highly non-Gaussian at lower
redshifts. The Gaussian shape of the PDFs at the beginning of
reionization, when the universe is almost completely neutral, is
driven by the primordial fluctuations in the density field of the
emitting IGM gas. At lower redshifts and near the end of reionization,
the completely ionized gas does not emit at 21-cm, while the
brightness temperatures of the leftover patches of neutral gas are still
governed by the density field inhomogeneities. At these redshifts,
entirely ionized cells contribute to the increasingly-dominant
delta-function at $T_b=0$~mK, while the emission from the
partially neutral cells maintains a Gaussian around a higher
$T_b$. The interplay between these two types of cells sets the shape
of the PDFs as the reionization proceeds.

In Figure~\ref{1pt}, we also plot the best fit of a Gaussian +
Exponential + (Dirac) Delta function model (GED model) for the 1-pt PDFs. This is an
``empirical'' model (i.e., based on simulation data), suggested by
\citet{ich09}:
\begin{equation}{p(T_b)} = \left\{ \begin{array}{cl}
   P_D\delta_D(T_b) + a\;{{\text{e}}^{\lambda T_b}}\ ; & T_b \leqslant
   {T_L}\ ,
\hfill  \\
   {c_G}\;{{\text{e}}^{ - \frac{{{{(T_b - {T_G})}^2}}}
{{2\sigma _G^2}}}}\ ; & T_b > {T_L}\ . \hfill  
 \end{array}  \right.
\end{equation}
To get a smooth curve, the values of the two functions and their first
derivatives are matched at the brightness temperature $T_b=T_L$,
leaving (after normalization) four independent parameters for the GED
model: the joining point of the exponential and the Gaussian function,
$T_L$, the mean of the Gaussian, $T_G$, its standard deviation,
$\sigma_G$, and its maximum value $c_G$.

\begin{figure}
\includegraphics[width=8.5cm,keepaspectratio=true]{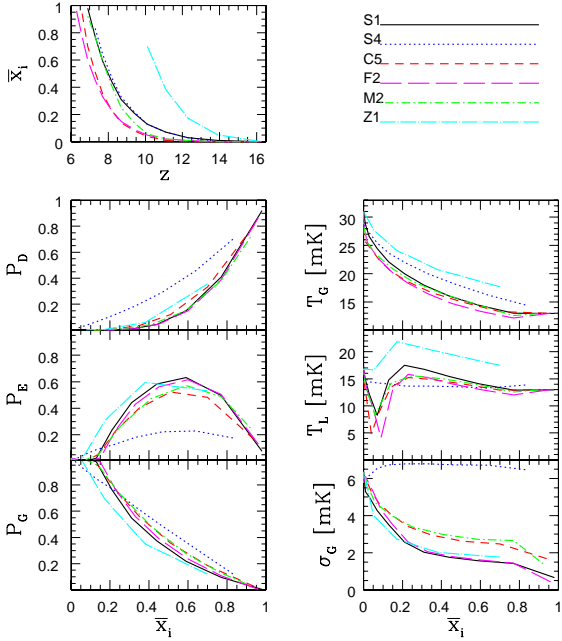}
\caption{Top panel: evolution of the global ionized fraction 
$\bar{x}_i$ with redshift $z$, for each of the six simulation runs.
Other panels show how various parameters derived from the
best-fit GED model to the 21-cm PDF evolve as the reionization proceeds. $P_D$, $P_E$, and $P_G$ are,
respectively, the fractional probability in the delta-function, the
exponential, and the Gaussian part of the PDF; $T_L$ is the joining
point of the exponential and the Gaussian, $T_G$ is the mean of the
Gaussian, and $\sigma_G$ is its variance. There are four
free parameters in the fit to each PDF.}
\label{params}
\end{figure}

From the best fit of the GED model to the PDF in each redshift slice
for each of the simulated data cubes, we obtain the value of the
probability fraction contained in the delta-function (at $T_b$'s
around zero), $P_D$, in the Gaussian (at high $T_b$'s), $P_G$, and in
the exponential part of the model (which interpolates between the
delta function and the Gaussian), $P_E$. These parameters can be
reconstructed from observations and must sum to unity, to ensure a
normalized PDF. The variation of $P_D$, $P_E$, $P_G$, $T_G$, $T_L$,
and $\sigma_G$ with the global ionized fraction, $\bar{x}_i$, is shown
in Figure~\ref{params}. The evolution of $\bar{x}_i$ with redshift for
each of the simulation runs is also shown.

%~~~~~~~~~~~~~~~~~~~~~~~~~~~~~~~~~~~~~~~~~~~~~~~~~~~~~~~~~~~~~~~~~~~~~~
%                                        FIGURE 2

\begin{figure}
\includegraphics[width=10cm,keepaspectratio=true]{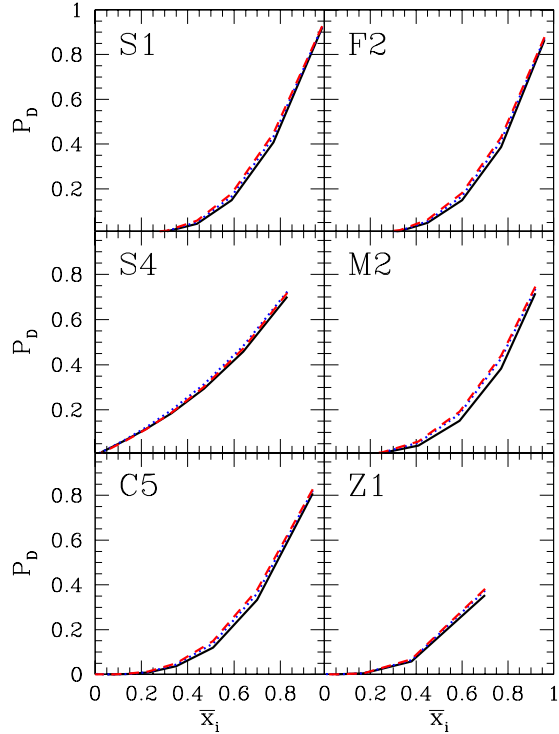}
\caption{Three different measurements of $P_D$ are shown 
versus $\bar{x}_i$. These include $P_D$ as calculated from the best
fit of the GED model (solid black curve), directly from the measured
PDF of $T_b$, i.e., the fraction of pixels that fall within the
lowest-temperature bin (dotted blue curve), and directly from the
measured PDF of $x_i$, i.e., the fraction of pixels that fall within
the highest-$x_i$ bin (dashed red curve). Results for the six different
simulation runs are shown.}
\label{c0_s1}
\end{figure}

When we fit the GED model, while its $\delta_D$ function portion is
meant to capture the PDF spike near $T_b=0$ mK (at low redshifts), we
do not attempt to model (or resolve in the data) the shape of this
spike. Thus, we exclude the lowest temperature bin before fitting the
GED model, and derive $P_D$ in three different ways: from the required overall normalization of the PDF to unity; then also directly, i.e., without the model fitting, from the total number count in the first bin of the
$T_b$ PDF; finally, we measure the one-point PDFs of $x_i$, and
estimate $P_D$ from the number count in the highest bin of this PDF (the
cells in this bin have $x_i \geq 0.95$). These three different
estimates of $P_D$ are compared for all six simulation runs in
Figure~\ref{c0_s1}. In this Figure, we show that the difference in
$P_D$ calculated from the cell counts and from the GED model is
negligible, which indicates that this model represents the data
faithfully. The values of $P_D$ as measured indirectly (from fitting
the GED model) yield an accurate estimate of the fraction of fully
ionized cells. In the limit of infinite resolution, $P_D$ would
equal $\bar{x}_i$ and so directly measure the reionization history,
while in reality $P_D$ measures a low-resolution, smoothed-out version
of the reionization history.

%~~~~~~~~~~~~~~~~~~~~~~~~~~~~~~~~~~~~~~~~~~~~~~~~~~~~~~~~~~~~~~~~~~~~~~~
%                                        FIGURE 3
 
Comparing the various simulation runs, we find that small-scale
structure has a relatively minor effect on the 21-cm PDF during
reionization, at least for the present implementation of sub-grid
astrophysics, and when the final reionization redshift is held
(relatively) fixed. Compared to our fiducial case (S1), we have three
simulations where mainly the small-scale structure has been adjusted:
a scenario with photoheating feedback (F2), one with evaporating
minihaloes (M2), and one with increased sub-grid clumping (C5). The
latter two also do not have merger-tree source
haloes. Figures~\ref{1pt} and \ref{params} show that these scenarios
have the effect of stretching out cosmic reionization, especially by
delaying its progress early on, when the rarity of high-mass haloes
makes feedback effective (F2), or the still-high density makes
recombinations important (C5), or the minihaloes have not yet
photo-evaporated (M2). However, in all these scenarios, the 21-cm PDF
at a given stage (as measured by $\bar{x}_i$) is fairly unchanged. In
particular, the evolution of the probabilities $P_D$, $P_E$, $P_G$,
and the parameters $T_G$ and $T_L$ is rather similar for these three
scenarios and S1. 

Strong changes in the properties of the ionizing sources do have more
of an effect on the evolution of the PDF. For example, a higher source luminosity
(Z1) leads to earlier reionization by somewhat more massive
haloes. Even early in reionization, the ionized bubbles are already
rather large (compared to the fixed pixel scale at which the PDF is
measured), which changes the PDF shape and the reconstructed
GED-model parameters. Also, since reionization in this case occurs at higher
redshifts, when the universe is denser, the mean 21-cm brightness
is higher, leading to higher values of $T_G$ and $T_L$ compared to the
lower-redshift cases. Furthermore, even without changing the redshift range, 
if the ionized sources lie in much more
massive haloes (S4) than in the S1 case, the impact on the PDFs is noticeable. 
In this case, the ionized bubbles (produced by
larger and more strongly-clustered sources) grow larger than the
effective PDF resolution quite early in reionization, so that $P_D$ is
larger than in the other cases (mostly at the expense of $P_E$), and
is generally closer to the value of $\bar{x}_i$.

In summary, we find that $\bar{x}_i$ is the main parameter determining
the one-point 21-cm PDFs. In particular, various modifications in the
small-scale structure have only a minor effect on the PDF evolution
versus $\bar{x}_i$ (as quantified by the parameters of the GED model
fits). This suggests that analysis of features of the observed
one-point PDFs can be used to reconstruct the global reionization
history relatively independently of any assumptions about the
astrophysics on unresolved sub-grid scales. On the other hand, the
typical mass of source haloes, and the typical reionization redshift,
have more of an effect. It is important to note, though, that
observations will provide independent constraints on these major
parameters. The redshift will obviously be measured, and, for example,
the span of the reionization epoch will constrain the typical halo
mass driving this process. Note also from Figure~\ref{params} that in
all the simulation runs, the Gaussian probability $P_G$ can be taken
as a rough estimate of the cosmic neutral fraction, $1-\bar{x}_i$.

While the one-point PDF is interesting, it will be rather difficult to
measure with the upcoming generation of instruments, mainly due to
comparatively bright foregrounds and the associated thermal noise. In
particular, \citet{ich09} found that the one-point PDF can only be
reconstructed from upcoming observations if the analysis is made on
the basis of quite strict (and not easily tested) assumptions that the
real PDF is very similar in shape to that measured in numerical
simulations. This difficulty motivates the use of an alternative
statistical tool that should have a much higher signal-to-noise ratio
for a given observation, namely the
\textit{difference PDF}\/ proposed by \citet{dif08}. In the next Section
we present the first numerical measurements of difference PDFs,
specifically using the same six simulated data cubes, and we discuss
their properties.

%%%%%%%%%%%%%%%%%%%%%%%%%%%%%%%%%%%%%%%%%%%%%%%%%%%%%%%%%%%%%%%%%%%%%%%
\subsection{Difference PDFs}\label{diff}

The PDF of the difference in the 21-cm brightness temperatures,
$\Delta T_b\equiv |T_{b1}-T_{b2}|$, at two separate points in the sky
(or, analogously, at two cells in the simulated data cube) was
suggested by \citet{dif08} as a useful statistic for describing the
tomography of the IGM during the EOR. More precisely, if we consider
two points separated by a distance $r$, then the distribution of
$\Delta T_b$ is given by the difference PDF $p_\Delta(\Delta T_b)$,
normalized as $\int p_\Delta \Delta T_b=1$. The motivation for
introducing this statistic is at least threefold. Firstly, the
effective number of data points available for reconstructing the
difference PDF is much larger than for the one-point PDF (by roughly a
factor of the number of pixels over two, though the pixel pairs must
be divided up into bins of distance $r$). Secondly, the difference PDF
(which is a major piece of the two-point PDF) is a generalization
which not only includes the information in the commonly considered
power spectrum or two-point correlation function (which can be derived
from the variance of the difference PDF), but also yields additional
information. And thirdly, being a PDF of a difference in $T_b$, it
avoids the mean sky background and fits naturally with the temperature
differences measured via interferometry.
%~~~~~~~~~~~~~~~~~~~~~~~~~~~~~~~~~~~~~~~~~~~~~~~~~~~~~~~~~~~~~~~~~~~~~~~
%                                        FIGURE 4
\begin{figure*}
\includegraphics[width=18cm,keepaspectratio=true]{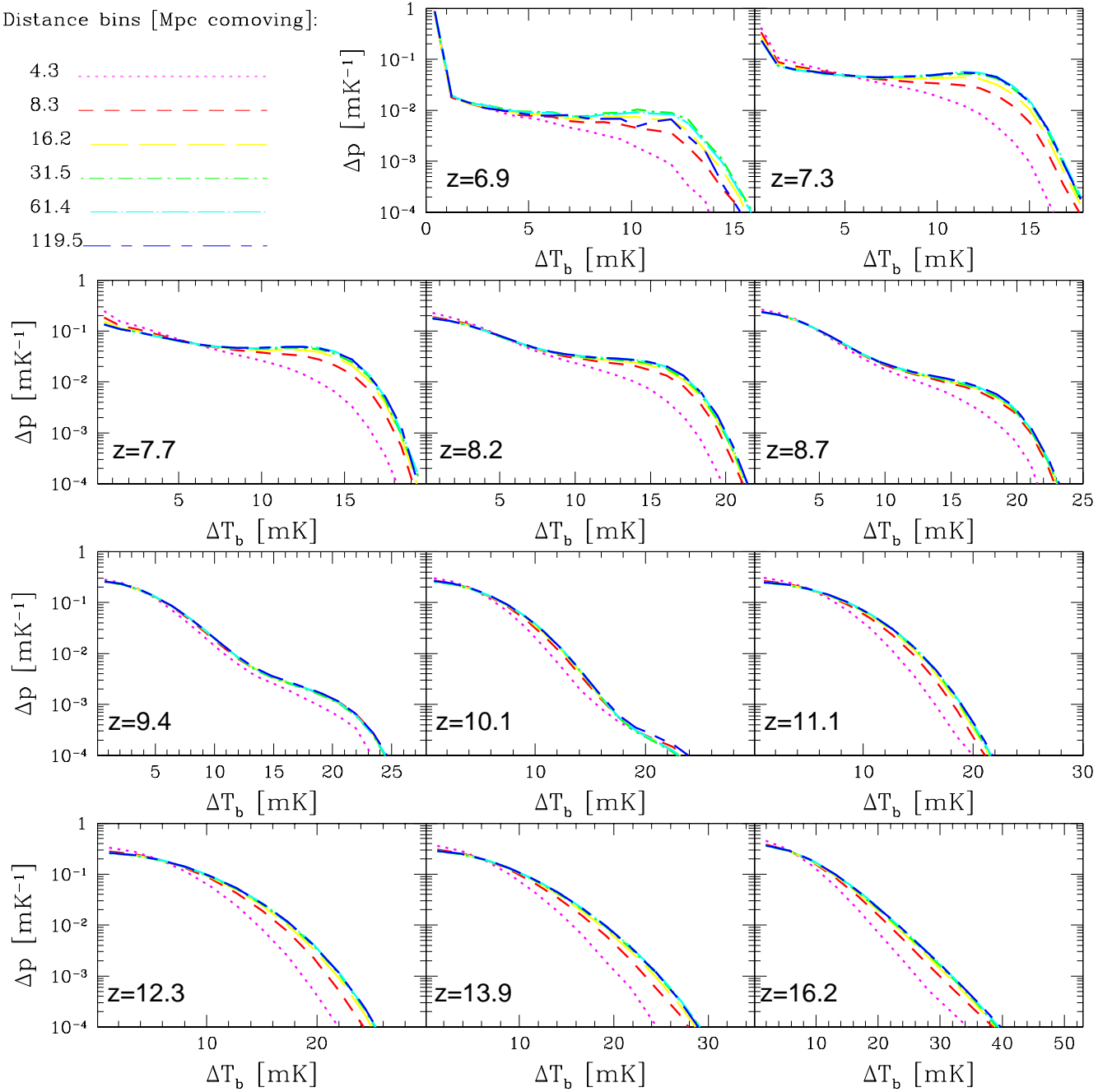}
\caption{Difference PDFs are shown for all the output redshifts of 
the $S1$ simulation. The $x$-axis is the brightness temperature
difference $\Delta T_b$ for pairs of cells, and the $y$-axis,
$p_\Delta$, is the number fraction of pairs of cells at a given
separation, $r$, with a given $\Delta T_b$ (normalized by the size of
the $\Delta T_b$ bin). Different curves indicate different $r$ bins
(the legend indicates the central values of
the logarithmic bins).\label{2pt_s1}}%
\end{figure*}

\begin{figure}
\includegraphics[width=16cm,keepaspectratio=true]{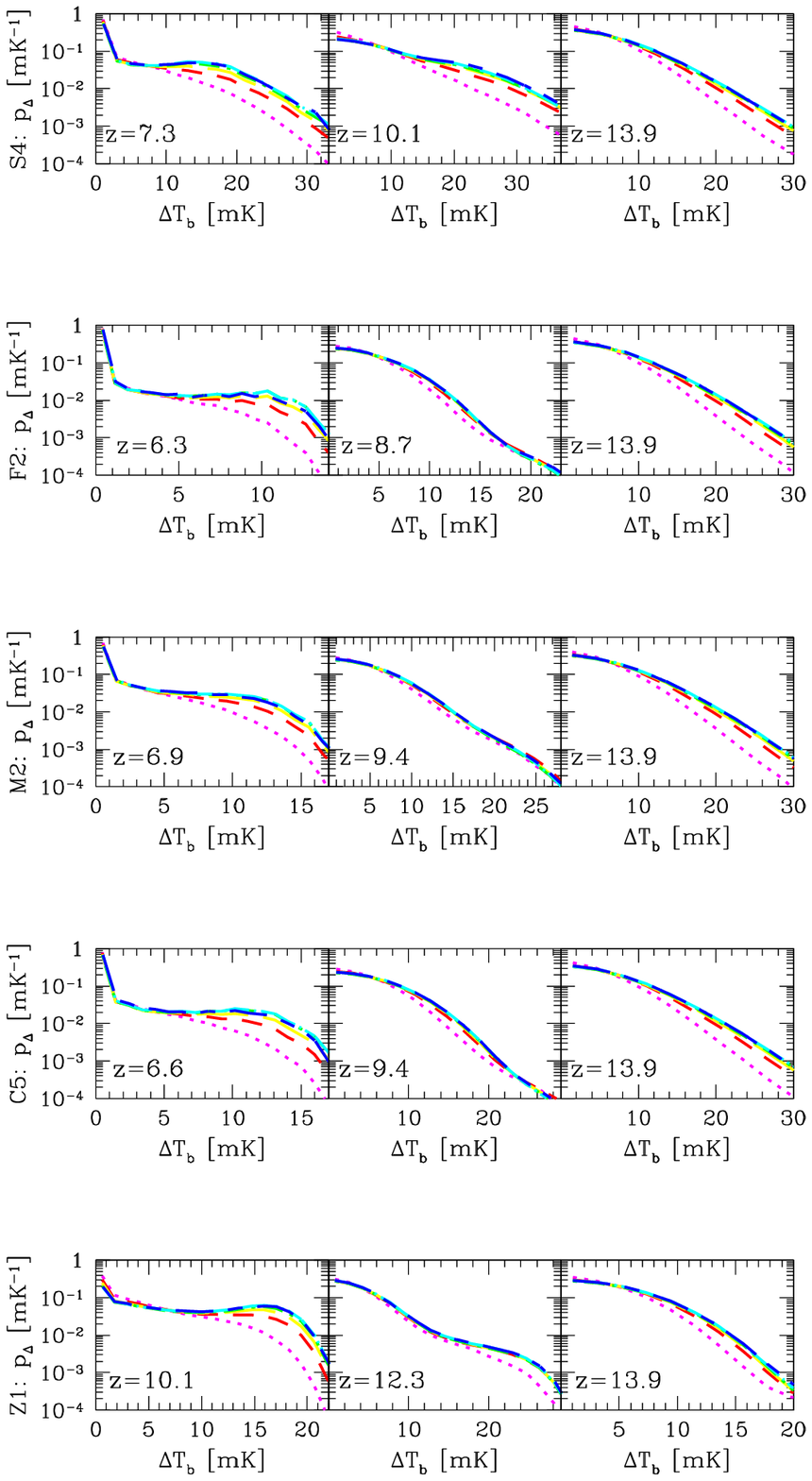}
\caption{This Figure is a smaller version of 
Figure~\ref{2pt_s1}: difference PDFs for the rest of the simulation
runs are shown for three representative redshifts (i.e., early, mid,
and late reionization). The legend is the same as in
Figure~\ref{2pt_s1}. \label{2pt}}%
\end{figure}

We present the first measurements of difference PDFs, for the same set
of \citet{mq07} simulation runs that we used to discuss the properties
of the one-point PDF in the previous Section. The difference PDFs for
all the redshifts of $S1$ are shown in Figure~\ref{2pt_s1}. For each
of the other five simulation runs, we show difference PDFs at three
representative redshifts in Figure~\ref{2pt}. In Figures~\ref{2pt_s1}
and \ref{2pt}, every redshift has $6$ distance bins, logarithmically
spaced, and each distance bin has $20$ temperature bins, linearly
spaced. The range in distance is chosen so that it covers basically
the full range from the resolution (pixel) scale to the largest
separations found within the 94 comoving Mpc data cube.                                       

Even for the one-point PDF, there is no good analytical model that
matches simulations, probably because the PDF is sensitive to the
reionization topology, specifically to the way in which the
complex-shaped ionized bubbles partially overlap the box-shaped
pixels. This led \citet{ich09} to base their analysis on the PDF as
measured in a simulation, and to consider an empirical model for
fitting the PDF shape. Similarly, in the case of the difference PDF,
the analytical model of \citet{dif08} does not quantitatively match
the result that we find in the simulations, but we can nonetheless use
the model and the discussion in \citet{dif08} to develop a qualitative
understanding of the difference PDF and how best to analyze it.

\begin{figure}
\includegraphics[width=10cm,keepaspectratio=true]{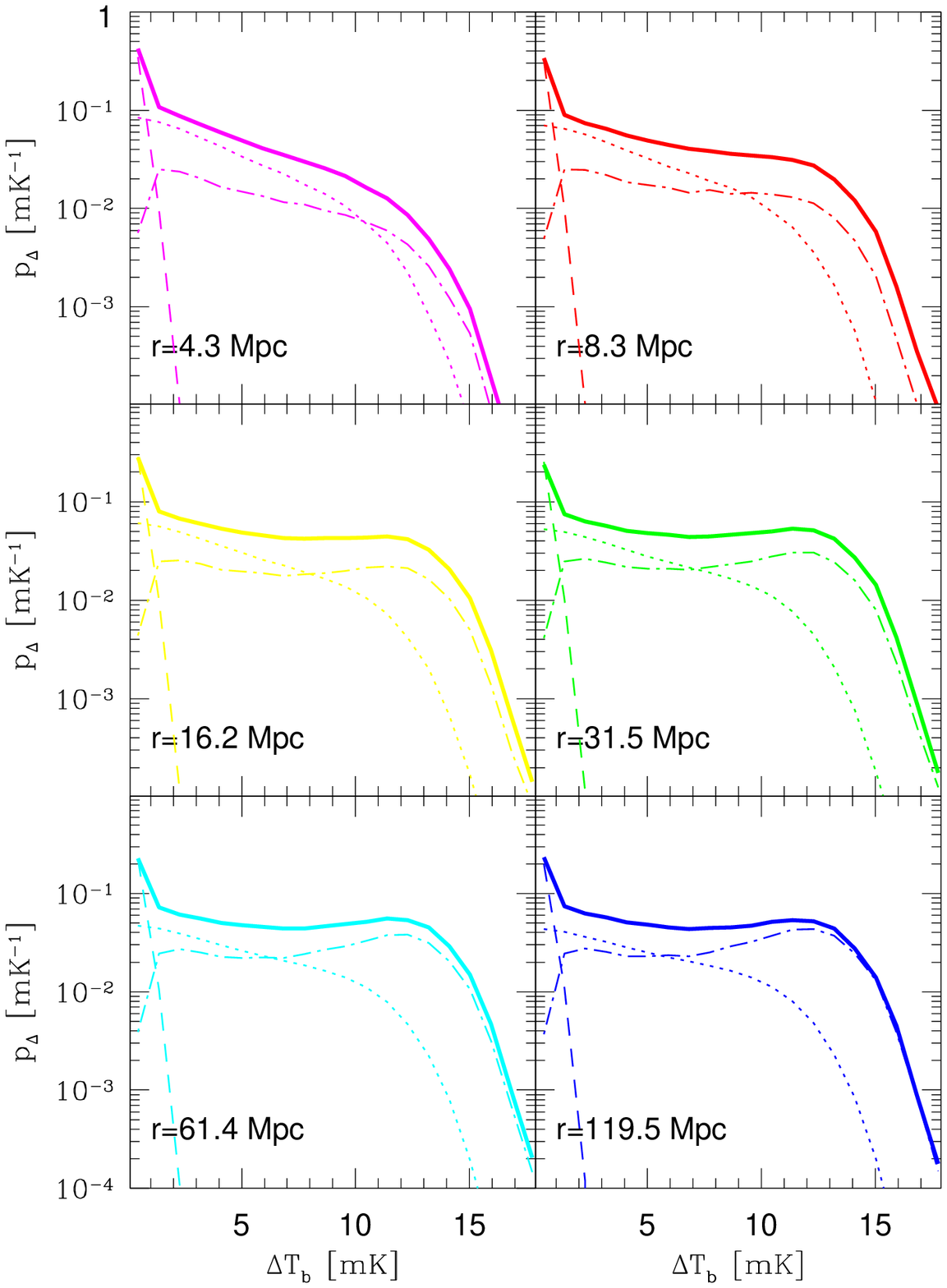}
\caption{Separate contributions to the difference PDF, for the
S1 simulation at $z=7.3$. We show the contribution of pixel
pairs in which both cells are fully ionized (dashed curves, capturing
the $\delta_D$ functions at zero $\Delta T_b$), pairs in which one cell is fully ionized
(dot-dashed curves), and pairs in which neither of the cells is fully ionized
(dotted curves). The total $p_\Delta$ (solid curves) is the sum
of these three contributions.}
\label{2pt_s1_sep}
\end{figure}

At high redshifts, when the PDF is nearly Gaussian, the difference PDF
(which is defined using the absolute value of $\Delta T_b$) should
approximately be a half-Gaussian. Non-linear density fluctuations,
though, give a slower decaying tail at high $\Delta T_b$ than would be
expected for a pure Gaussian. As reionization develops with time, the
difference PDF becomes a superposition of three contributing terms:
The pixel pairs in which both pixels are fully ionized, those in which
one pixel is (partly) neutral and the other ionized, and those where
both are neutral. We explicitly show these three contributions in
Figure~\ref{2pt_s1_sep} for one redshift near the midpoint of
reionization in the $S1$ simulation. The both-ionized pixel pairs
basically give the $\delta_D$ function at zero $\Delta T_b$; the
amount of probability in this $\delta_D$ function is physically
meaningful, as discussed below. The pairs in which one pixel is
neutral and the other ionized tend to be well separated in $T_b$, and
so this contribution is responsible for the high $\Delta T_b$ peak, at
$\Delta T_b \sim 13$~mK in the case plotted here. As we consider
smaller $r$, the $T_b$ values of the two points that are separated by
$r$ become more strongly correlated, making it difficult for one to be
ionized and the other neutral, and so this contribution declines at
smaller $r$. At the same time, as we reduce $r$, this contribution
becomes more highly concentrated at small values of $\Delta T_b$,
since at small separations, if one pixel is fully ionized, then the
second one tends to be at least highly ionized, making their $T_b$
difference small.  Note also that the ionized+neutral contribution
drops suddenly at $\Delta T_b \sim 1.5$~mK, but this is due to the
fact that some of the probability in this region that should be
included under ionized+neutral is incorrectly swept up under the
both-ionized $\delta_D$ function, which is dominant at $\Delta T_b$
near zero. This is an unavoidable effect of the finite (2.9 Mpc)
resolution of our gridded field. Since in practice we define a ``fully
ionized'' pixel as having an ionization fraction of $95\%$ or higher,
some highly ionized pixels are classified as fully ionized; a higher
resolution map, with a definition closer to $100\%$ ionization, would
move this artificial drop-off closer to $\Delta T_b=0$. Those pairs in
which both pixels are neutral peak at $\Delta T_b=0$, and give a
contribution with a roughly half-Gaussian shape, at all separations
$r$.
%~~~~~~~~~~~~~~~~~~~~~~~~~~~~~~~~~~~~~~~~~~~~~~~~~~~~~~~~~~~~~~~~~~~~~~~
%                                        FIGURE 5

%~~~~~~~~~~~~~~~~~~~~~~~~~~~~~~~~~~~~~~~~~~~~~~~~~~~~~~~~~~~~~~~~~~~~~~
%                                        FIGURE 6

The difference PDF as a function of separation transitions between two
limits. The $r \to \infty$ limit corresponds to two
uncorrelated points, for which $p_\Delta$ is essentially a convolution
of the one-point PDF $p$ with itself. As long as $r$ is large enough
to maintain a weak correlation, $p_\Delta$ keeps its
large-$r$ limit and only varies slowly as $r$ is decreased. Once $r$
becomes small enough for a significant correlation (which is positive
in the physical regimes considered here), it becomes harder to produce
a large difference $\Delta T_b$ between the two correlated points (as
noted above, even when one is fully ionized and the other is not);
$p_\Delta$ thus becomes more strongly concentrated near $\Delta T_b =
0$, approaching a $\delta_D$ function at $\Delta T_b = 0$ as $r
\to 0$.

The correlation between cells, referred to above, probes different
physical effects at different redshifts. Before reionization, this is
the density correlation, which arises from large-scale modes in the
initial fluctuations. During reionization, the correlation of $T_b$ is
dominated by ionization, so that points close enough together to be in
the same ionized bubble (or in strongly correlated nearby bubbles)
will have strongly correlated 21-cm brightness temperatures.  Thus, by
inspecting the plots, one can make a rough estimate of the average
size of an ionized bubble at low redshifts, or the typical density
fluctuation correlation length at high redshifts. This effective
correlation length is the first separation bin at which the difference
PDF at high $\Delta T_b$ drops significantly below its value at larger
separations. For example, in Figure~\ref{2pt_s1}, the average bubble
size can be seen to increase beyond 10 comoving Mpc during the late
stages of reionization.

\begin{figure}
\includegraphics[width=8cm,keepaspectratio=true]{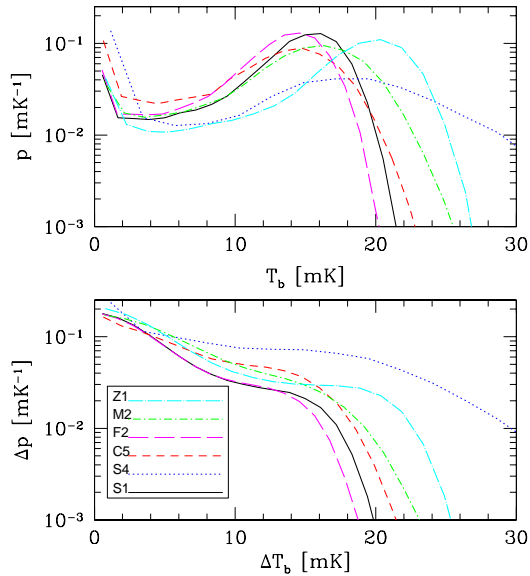}
\caption{Comparison of the one-point PDFs (upper panel) and 
difference PDFs (lower panel), for all six simulation runs, at $r=16.2$ comoving Mpc. Each
simulation run is shown at the redshift for which the value of the
global ionized fraction $\bar{x}_i$ is closest to $0.4$, i.e. in the
midst of the reionization process.}
\label{2pt_comparison}
\end{figure}

As in the case of the one-point PDF, the difference PDF is relatively
insensitive to variations in the small-scale sub-grid physics, as
tested by the various simulation runs. Figure~\ref{2pt_comparison}
displays a comparison of the difference PDFs for the six different
reionization runs, for various comoving distance bins, at the redshift
where $\bar{x}_i$ is closest to the value of $0.4$ (i.e., in the midst
of the reionization process). We also show the corresponding one-point
PDFs, for easy comparison. Similarly to $p$ (as discussed in the
previous section), the large ionized bubbles and correlation length in
the case of reionization by massive, rare sources stretch $p_\Delta$
out to higher values of $\Delta T_b$ (as seen in the Z1 run and,
especially, S4). Our findings are consistent with those of
\citet{mq07} and with the general theoretical 
expectation that clustered groups of galaxies determine the spatial distribution of ionized bubbles \citep{bl04} which is then driven
by large-scale modes and is mainly sensitive to the overall bias of
the ionizing sources.

%~~~~~~~~~~~~~~~~~~~~~~~~~~~~~~~~~~~~~~~~~~~~~~~~~~~~~~~~~~~~~~~~~~~~~
%                                        FIGURE 7

We next proceed to measure the parameter analogous to $P_D$, but this
time for the case of difference PDFs. This parameter, which we denote
$\Delta P_D$, represents the (number) fraction of pairs for which
$\Delta T_b \approx 0$. In the reality of having limited resolution, it is the
fraction of pairs for which $\Delta T_b$ falls within the lowest temperature bin.
Ideally, this value would directly measure the fraction of pairs in
which both cells are fully ionized, but the finite resolution adds 
 a contribution from pairs that do not satisfy this condition, but
nonetheless have matching brightness temperatures to within the bin
size. 

\begin{figure}
\includegraphics[width=11cm,keepaspectratio=true]{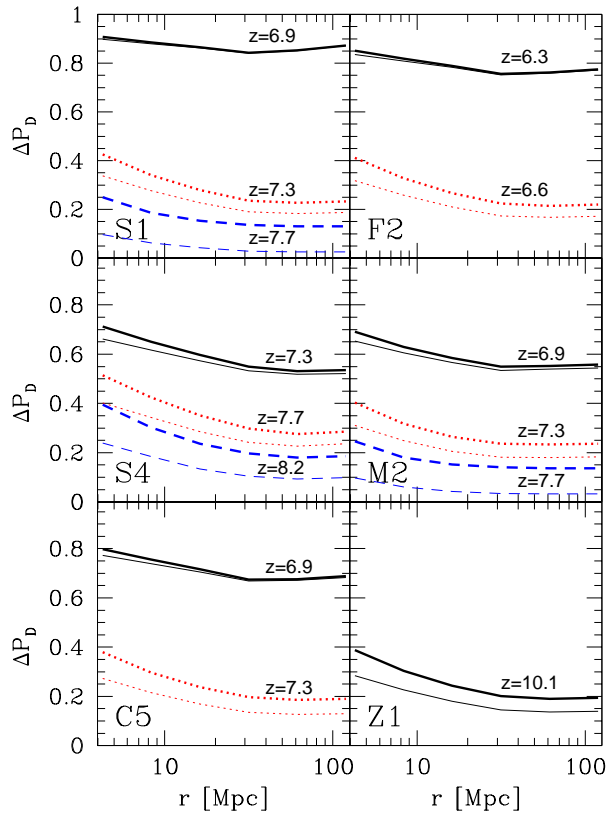}
\caption{$\Delta P_D$ parameter, measured from difference PDFs 
as the value of the zero-temperature-difference bin, for the S1
simulation, is shown for the redshifts where it is non-vanishing. It
represents the fractional probability in the delta function around
$\Delta T_b\approx 0$, i.e. the number fraction of pairs of cells
which are either both fully ionized, or both at the same brightness
temperature. We see that this value tends to asymptote to the
corresponding $P_D$, at $r\sim 0$, as expected for fully correlated
cells. The thin lines of the same type and color represent the part of
the corresponding $\Delta P_D$ that originates from pairs where both
cells are fully ionized.}
\label{2ptc0}
\end{figure}

Just as $P_D$ in the one-point PDF measures a low-resolution version
of the reionization history, so can $\Delta P_D$ be considered as
measuring a low-resolution version of the ionization correlation
function\citep{dif08}. In particular, in the limit of infinitely high
resolution, $\Delta P_D$ would precisely measure the joint ionization
probability of two points as a function of their separation. In
this case, we would expect $\Delta P_D$ to vary from the corresponding
value of $P_D$, at $r\to 0$, down to $P_D^2$, at $r\to \infty$ (where
each pixel in the pair is independently ionized with probability
$P_D$). While these relations are not exact with finite resolution,
they do provide a rough guide for what to expect. We show the value of
$\Delta P_D$ in Figure~\ref{2ptc0}, for the redshifts of each
simulation at which the $\delta_D$ function at $\Delta T_b \approx 0$ is
visible. A comparison with the corresponding values of $P_D$ (shown in
Figure~\ref{c0_s1}) shows that the above theoretical behavior is
satisfied only approximately, since the fully ionized pairs make up
only a fraction of $\Delta P_D$. Still, the Figure shows the flat
asymptote of $\Delta P_D$ at large $r$, and its rise as $r$ drops
below the correlation length (although $r$ does not quite reach small enough
values to see the flat $r\to 0$ asymptote). 

%~~~~~~~~~~~~~~~~~~~~~~~~~~~~~~~~~~~~~~~~~~~~~~~~~~~~~~~~~~~~~~~~~~~~~
%                                        FIGURE 8

We have illustrated how the difference PDF encodes information about
the EOR, in particular by separating out information on the
ionization correlations (unlike the power spectrum analysis, in which the
ionization and density correlations are mixed together). We leave for
future work a more quantitative analysis of the features of the
difference PDF and their relation to the properties of the ionizing
sources and the reionization history.

\section{Summary and Conclusions}\label{conclude}

Upcoming low-frequency radio observations will use the MWA, LOFAR, and
similar instruments to survey the sky for redshifted 21-cm emission. It
is therefore important to develop statistical tools that can be used
to extract information about the history of reionization from these
observations. In this paper, we examine PDFs and difference PDFs,
using six different simulations from \citet{mq07}. We show that
the PDFs are highly non-Gaussian in the midst of the
reionization process, and are thus a complementary statistic to the
commonly discussed power spectrum. As a way to analyze the PDFs, we
examine the evolution of the parameters of the best-fit GED model with the global ionized fraction $\bar{x}_i$.
 In particular, the
$\delta_D$ function portion of the probability ($P_D$) measures a low-resolution,
smoothed-out version of the reionization history (i.e., $\bar{x}_i$
as a function of redshift).

We also present the first numerical measurements of difference PDFs; specifically, we measure difference PDFs for the same set of simulation runs from \citet{mq07}. We argue that the larger data set and the nature of this
statistic can be significant advantages in the presence of bright
foregrounds. The difference PDF can be physically understood as
arising from three contributions: pixel pairs in which both, one, or
neither of the pixels is ionized. As an illustration of the information that can be
deduced from the difference PDFs, we consider the typical
correlation length, which corresponds to the average size of an
ionized bubble during reionization, or the typical density fluctuation
correlation length, at the onset of the EOR. The difference PDF also has a
delta-function portion ($\Delta P_D$), which measures a
low-resolution, smoothed-out version of the ionization correlation
function at each redshift.

We find that increasing small-scale clumping, and including photoheating feedback
or minihaloes has only a small effect on the one-point and difference
PDFs (considered at a given $\bar{x}_i$), at least
within the range of assumptions covered by the simulations that we
considered. On the other hand, the PDFs are highly sensitive to the
properties of the ionizing sources, so that measuring them can help
distinguish between reionization driven by large versus small haloes
and help us unveil information about the first sources of light in the
universe. These conclusions parallel those of \citet{mq07}, highlighting the fundamental fact that the spatial
structure of reionization is driven by large-scale modes and depends
mainly on the overall bias of the ionizing sources \citep{bl04,fzh04}.

We plan to more precisely quantify the properties of difference PDF
and establish the relation between their features and the properties
of the IGM during the EOR. It would also be interesting to
explore the PDFs and difference PDFs in alternative reionization
scenarios, such as those dominated by X-ray sources or a decaying dark
matter particle. Contrasting different scenarios may lead to a fuller
understanding of the information content of the specific PDF shapes that we measured.

%%%%%%%%%%%%%%%%%%%%%%%%%%%%%%%%%%%%%%%%%%%%%%%%%%%%%%%%%%%%%%%%%%%%%%%

\section*{Acknowledgments}

We thank Matthew McQuinn for useful comments, and we are grateful for the hospitality of the Aspen Center for Physics, where
part of this work was completed. This work was supported (for VG) by DoE DE-FG03-92-ER40701, NASA NNX10AD04G, and
the Gordon and Betty Moore Foundation, and (for RB) by the Moore
Distinguished Scholar program at Caltech, the John Simon Guggenheim Memorial Foundation, and Israel Science
Foundation grant 823/09.

%\bsp

\label{lastpage}

\end{document}